\documentstyle[12pt]{article}
\newcommand{\sect}[1]{\setcounter{equation}{0}\section{#1}}

\begin{document} \def\bq{\begin{equation}}
\def\eq{\end{equation}}
\begin{flushright}
LPTENS/98/28
\end{flushright}

\begin{center}
{\large\bf
Dynamics versus replicas in  the random field Ising model }
\end{center}

\begin{center}
{\bf E. Br\'ezin$^{a)}$ and C. De Dominicis$^{b)}$} \end{center} \vskip 2mm
\begin{center}{$^{a)}$ Laboratoire de Physique Th\'eorique, Ecole Normale
Sup\'erieure}\\ {24 rue Lhomond 75231, Paris Cedex 05,
 France{\footnote{ Unit\'e propre 701 du
Centre national de la recherche scientifique, associ\'ee \`a l'Ecole
Normale Sup\'erieure et \`a l'Universit\'e de Paris-Sud} } }\\
{$^{b)}$ Service de physique th\'eorique}\\ {Saclay, 91190 Gif-sur-Yvette,
France}\\
\end{center}
\vskip 3mm
\begin{abstract}
 In a previous article we have shown, within the replica formalism, that
the conventional picture of the random
field Ising model breaks down, by the effect of singularities in the
interactions between fields involving
several replicas, below dimension eight. In the
zero-replica limit several coupling constants have thus to be considered,
instead of just one. As a result we had found that there is no stable fixed
point in the
vicinity of dimension six. It is natural to reconsider the problem in a
dynamical framework, which does not require replicas, although
the equilibrium properties should be recovered in the large time limit.
Singularities in the
zero-replica limit are a priori not visible in a dynamical picture. In this
note we show that in fact new interactions are also
generated in the stochastic approach. Similarly these interactions are
found to be singular below dimension eight. These critical singularities
require the introduction of a time origin $t_0$ at which initial data are
given. The dynamical properties
are thus dependent upon the waiting time. It is shown here that one can
indeed find a complete correspondence between the
equilibrium singularities in the
$n=0$ limit, and the singularities in the dynamics when the initial time
$t_0$  goes to minus infinity, with $n$ replaced by $-\frac{1}{t_0}$. There
is thus complete coherence between the two approaches.
\end{abstract}
\newpage

\vskip 5mm \sect{Introduction}

In a recent article \cite {DD-EB} we have reconsidered the old problem of
the random field Ising model (RFIM). Using the replica approach, we started
with
a $\phi^4$ field theory, but included the five marginal operators of  same
dimension, namely $g_1\sum\phi_a^4, g_2\sum\phi_a^3\sum\phi_b,
g_3(\sum\phi_a^2)
^2, g_4\sum\phi_a^2(\sum\phi_b)^2, g_5(\sum\phi_a)^4$, in which the sums
run from $1$ to $n$. The last four operators are normally not considered,
since
they are a priori of relative order, $n, n^2$ or $n^3$ with respect to the
first operator. Within this  naive $n=0$ limit one recovers the usual
correspondence,
between the RFIM in dimension $d$ and the pure system in dimension $(d-2)$
\cite{AIM,Young, Parisi-Sourlas}, which is known to break down in three
dimensions \cite{Imbrie,Bricmont}.\\

However we have shown in \cite {DD-EB} that the statistical fluctuations
give singular contributions to the effective interactions
 between
several replicas
$g_i^{eff}$, with $2\leq i\leq5$. They are singular below dimension eight
in the critical region when $n$ goes to zero, invalidating
thus the naive zero-replica limit. For instance $g_3^{eff}$ receives a
one-loop contribution due to $g_1$ which is proportional
(at zero momentum) to
\bq I_n(r) = g_1^2\int d^{d}q \ \frac{1}{(q^2+r)^2}\
\frac{\Delta^2}{({q^2}+r-n\Delta)^{2}}\eq
in which $\Delta$ is the variance of the Gaussian random field, and $r$ is
proportional to the distance to the critical temperature. In the critical
region
in which $r$, which goes to zero, is of order $n\Delta$,  $I_n(n\Delta)$
diverges for $n$ small in dimensions lower than eight as
\bq I_n(n\Delta) \sim \frac{1}{n^{(8-d)/2}}. \eq
As a result $g_3^{eff}$  is of order $1/n$ in dimension six and this
invalidates the simple $n=0$ limit. Consequently we have studied the
renormalization group
flow  in dimension $6-\epsilon$ and found that the usual dimensional
reduction fixed point is unstable, as well as all the other fixed points
allowed
by this flow  \cite {DD-EB}.

In this work we want to reconsider the same problem, within the dynamical
approach which is free of replicas, and is then physically
more transparent. Formally, J. Kurchan \cite {Kurchan} has shown, within
the supersymmetric approach  \cite {Zinn}, that
one gets exactly the same action with fields bearing replica indices (for
the statics) and with  fields bearing time-indices
(in the dynamics), except for the time derivative term. However the
detailed relationship between the two formalisms is far from clear,
especially whenever
singularities associated with the number of replicas are involved. Since
equilibrium is the large time limit of a stochastic process, the above
mechanism ought to
be understood as well in a dynamical framework
\cite{CDD}.
We have thus
considered a Langevin stochastic equation, and the  corresponding field
theory obtained through the averaging over the  noise and over the random
field. In this
approach we find also singularities below dimension eight related to new
effective dynamic interactions. In order to cope with these singularities,
we
have to introduce a
"waiting" time, i.e. we assume that we start with some initial data at
finite time $t_0$. The singularities manifest themselves then as
divergences when $t_0$
goes to minus infinity, with a correspondence between $-t_0$ and $1/n$. We
believe that this supports significantly the results of the analysis that
we had
developped within the replica approach \cite {DD-EB}.
\sect{ The Langevin formalism} 
We consider a simple Langevin dynamics for a
 $\phi^4$ theory,
\bq \frac{\partial{\phi(x,t)}}{\partial {t}} = - \frac{\delta S}{\delta
\phi} + h(x) + \eta(x,t),\eq
 with a white noise source
\bq <\eta(x,t) \eta(x',t')> = \gamma \delta(t-t') \delta^{(d)} (x-x'), \eq
 a time-independent random field,
\bq \overline{h(x)}= 0\  ,\  \overline{h(x) h(y)} = \Delta
\delta^{(d)}(x-y).\eq
 for an action
	\bq
 S(\phi) =\int dx [{1\over{2}} (\nabla\phi)^2 +{1\over{2}} r_0 \phi^2 +
{g\over{4!}}
\phi^4 ] . \eq

From this, one derives easily the bare (i.e. $g=0$) correlator defined as
\bq C(t,t',p; t_0) = \int d^{d}x \exp{(ip\cdot x)}\ \overline
{<\phi(x,t)\phi(0,t')>} \eq
with initial data
\bq \phi(x, t_0) = 0.\eq
The result is
\bq\label{corr} C(t,t',p; t_0) = \frac{\gamma}{2E_p} [ e^{-E_p\vert
t-t'\vert} - e^{-E_p(t+t'-2t_0)} ] +
\frac{\Delta}{E_p^2} [1- e^{-E_p( t-t_0)}] [1- e^{-E_p( t'-t_0)}] \eq
with
\bq E_p = p^2 + r_0 \  .\eq
If we let the initial time $t_0$ go to minus infinity, we recover the
time-translation invariant correlator
\bq \label{correl} C(t,t',p; -\infty) = \frac{\gamma}{2E_p}  e^{-E_p\vert
t-t'\vert}  +
\frac{\Delta}{E_p^2}\   , \eq
and its Fourier transform with respect to $(t-t')$ is
\bq \tilde {C} (\omega, p) = \frac {\gamma}{2\pi}\
\frac{1}{(p^2+r_0)^2+\omega^2 } + \frac{\Delta}{(p^2+r_0)^2}
\delta(\omega). \eq

The field theory related to this stochastic process is derived in the
standard way \cite{Martin, Zinn}:
\begin{eqnarray} \overline {<\it{O}(\phi)>} &=& \int D\phi \it{O}(\phi)
\det(\frac{\partial}{\partial {t}}+\frac{\delta^2 S}{\delta \phi^2})
\overline{<\prod_{x,t} \delta(\frac{\partial{\phi(x,t)}}{\partial {t}} +
\frac{\delta S}{\delta \phi}
 - h(x)-\eta(x,t))>} \nonumber\\&=& \int D\phi D\hat{\phi}DcD\bar{c}
\it{O}(\phi) \exp{-\int dxdt \bar{c}\left(\frac{\partial}{\partial
{t}}+\frac{\delta^2
S}{\delta \phi^2}\right)c}
\nonumber\\
&&\overline{<\exp i\int dxdt \hat{\phi}
(\frac{\partial{\phi(x,t)}}{\partial {t}} +
\frac{\delta S}{\delta \phi}
 - h(x)-\eta(x,t))>} \nonumber\\
&=&\int D\phi D\hat{\phi}DcD\bar{c} \ \it{O}(\phi)\  \exp{-\int dxdt
\bar{c}\left(\frac{\partial}{\partial {t}}+\frac{\delta^2
S}{\delta \phi^2}\right)c}
\nonumber\\
&&\exp {i\int dxdt \hat{\phi} (\frac{\partial{\phi(x,t)}}{\partial {t}} +
\frac{\delta S}{\delta \phi}})\nonumber\\
 && \exp{-\frac{\gamma}{2} \int dxdt \hat{\phi(x,t)}^2 -\frac{\Delta}{2}
\int dx \int dt \int dt'\hat{\phi(x,t)}\hat{\phi(x,t')}} .
\end{eqnarray}
This is summarized in an effective action
\begin{eqnarray}\label{effaction} S_{eff} &=& \int dxdt \
\bar{c}\left(\frac{\partial}{\partial {t}}+\frac{\delta^2
S}{\delta \phi^2}\right)c -i\int dxdt \hat{\phi}(x,t)
\left(\frac{\partial{\phi(x,t)}}{\partial {t}} +
\frac{\delta S}{\delta \phi}\right)\nonumber\\&+&\frac{\gamma}{2} \int dxdt
\hat{\phi}^2(x,t) +\frac{\Delta}{2} \int dx \int dt \int
dt'\hat{\phi}(x,t)\hat{\phi}(x,t') ,\end{eqnarray}
in which $c(x,t)$ and $\bar{c}(x,t)$ are Grassmanian fields.\\

The quadratic part of this effective action allows one to recover the
$\phi\phi$ correlator (\ref{corr}), as well as the response
\bq \label{response}R(p,t,t') = \int d^{d}x <\phi(x,t)\hat{\phi}(0,t')> =
i\theta(t-t') \exp{-\left( E_p(t-t')\right)}.\eq
Note that the effective action (\ref{effaction}) contains only one
interaction term, local in time, proportional to
$\hat{\phi}(x,t)\phi^3(x,t)$.

\sect{Singularities below dimension eight}

Let us assume first that we let the initial time $t_0$ go to minus
infinity, and work with the simple correlator (\ref{corr}). We consider the
lowest order contribution to the four-point function
$\phi\hat{\phi}\phi\hat{\phi}$ at zero-momentum, zero frequency. It
consists of
a simple one-loop graph, which involves the integral of the product of two
correlators (\ref{correl}). The random-field part
of the correlator (\ref{correl}) gives a contribution proportional to
$g^2\Delta^2 \int d^{d}p \frac{1}{(p^2+r)^4}$ which diverges in the
critical domain for $d<8$. (Note that in the static replica approach one
has an identical critical singularity  for the zero-momentum
$\phi_a^2\phi_b^2$ function.)  \\

In order to deal properly  with this singularity we keep a finite initial
time $t_0$ and compute again the lowest order
contribution to the
four-point function $\phi(t)\hat{\phi}(t)\phi(t')\hat{\phi}(t')$ at
zero-momentum. It consists of an integral over the
momenta only of the
product of two correlators (\ref{corr}). Keeping only the $\Delta^2$ part
we have to consider
\bq\label{graph} I(t,t';t_0) =g^2\Delta^2 \int d^{d}p \frac{1}{E_p^4} [1-
e^{-E_p( t-t_0)}]^2 [1- e^{-E_p( t'-t_0)}]^2  , \eq
with $E_p = p^2 +r$. If we let $t_0$ go to $-\infty$, we recover the
previous critical singularity in dimensions
lower than eight as it should. If we keep $t_0$ finite, but let it grow to
large negative values, we have to integrate out
(\ref{graph}) over $p$. Using the identity
\bq  \frac{1}{E} [1- e^{-E( t-t_0)}] = \int_{0}^{t-t_0} d\tau e^{ -E\tau}, \eq
we end up after a few steps with
\bq\label{(3.3)} I(t,t';t_0) \sim (-t_o)^{(8-d)/2} f(u), \eq
in which $f(u)$ is the function of the single variable
\bq  u = (-t_0)r\eq
given by
\bq f(u) = u^{\frac{d-8}{2}}\int_{0}^{\infty}dx
\frac{x^{\frac{d}{2}-1}}{(1+x)^4} [1-e^{u(1+x)}]^4. \eq
Therefore for finite, but large, negative $t_0$, when one enters in the
critical region with $r$ of order $1/\vert t_0\vert$
the four-point function $\phi(t)\hat{\phi}(t)\phi(t')\hat{\phi}(t')$
receives contributions of order $\vert t_o\vert^{(8-d)/2}$
independent of $t$ and $t'$. A new interaction vertex, proportional
to $\vert t_o\vert^{(8-d)/2}\int dx\int
dtdt'\phi(x,t)\hat{\phi}(x,t)\phi(x,t')\hat{\phi}(x,t')$,  not present in
the initial Langevin dynamics
is thus generated, in very much the same way that in the replica approach a
vertex of the form
$n^{(8-d)/2}\int dx \left[\sum_{a} \phi_a^2(x)\right]^2$ resulted from the
singularity in less than eight dimensions.\\

A similar analysis of the dynamical fluctuations at two-loop order lead to
three more singular  contributions with four external legs. They are
respectively
proportional to \\
$\mid t_o\mid^{(8-d)/2}\int dx\int
dtdt'dt''\phi(x,t)\hat{\phi}(x,t)\phi(x,t')\phi(x,t'')\theta(t-t')$,\\
$\mid t_o\mid ^{(8-d)}\int dx\int
dtdt'dt''\phi(x,t)\hat{\phi}(x,t)\hat{\phi}(x,t')\hat{\phi}(x,t'')$,\\
 $\mid t_o\mid ^{\frac{3(8-d)}{2}}\int dx\left(\int
dt\hat{\phi}(x,t)\right)^4$.\\

We have thus a total of five dynamical interactions of degree four in the
operators $\phi$ and $\hat{\phi}$.
This is similar to the five operators that we had to
consider  in the replica approach, namely $\sum_a \phi_a^4,\
n^{-\frac{8-d}{2}}(\sum_a\phi_a^2)^2,\ \
n^{-\frac{8-d}{2}}(\sum_a\phi_a^3)\sum_b\phi_b,
 \\ n^{-(8-d)}(\sum_a\phi_a^2)\left(\sum_b\phi_b\right)^2,\
n^{-\frac{3(8-d)}{2}}(\sum_a\phi_a)^4$. Clearly $1/n$ is replaced in the
dynamics
by $\vert t_0\vert$.

It is interesting to analyse the renormalization group flow with those five
operators. Some diagrams, which would vanish in the replica
approach because they have an explicit relative factor of $n$, in spite of
the singular behavior of the interactions, are identically zero in the
dynamics by
causality; this is due to the retarded nature of the response functions which
forbids to make closed loops by propagating only response functions.
Whenever the singular behavior leads to a finite result in the $n=0$ limit
of
the replica approach, the singular dependence in $t_0$ does lead to the
same conclusion in the dynamics.

\sect{Conclusion}
 We have shown that the dynamical approach to the dynamics of a random
field Ising model (or $\phi^4$ theory), shows that new
singular interactions, absent from the bare dynamical action corresponding
to the Langevin equation,  are generated
by the fluctuations  below dimension eight.These new interactions cannot be
easily cast into
a modified Langevin equation. These interactions have singularities in the
waiting time, analogous
to the singularities that we had found earlier in the replica approach,
with a simple correspondence between $1/n$ and $\vert t_0\vert$.
The conclusions that we had drawn from our earlier work, namely that in the
space of the five coupling constants there was no stable fixed point
of order $\epsilon = 6-d$, are thus entirely  supported by the dynamics.

\begin{center}
{\bf Acknowledgement}
\end{center}
It is a pleasure to acknowledge discussions with  L. Cugliandolo and  M.
M\'ezard
during the course of this work. We would like to thank the anonymous
referee of our previous work, for raising the question of the dynamics
which gave rise to this note.



\begin{thebibliography}{99}
\bibitem{DD-EB} E. Br\'ezin and C. De Dominicis, cond-mat 9804266.
\bibitem{AIM} A. Aharony,Y.Imry and S.K. Ma, Phys. Rev.Lett. {\bf 37}
(1976) 1364.
\bibitem{Young} A.P.Young, Phys. Rev.Lett. {\bf 37} (1976) 944.
\bibitem{Parisi-Sourlas} G.Parisi and N.Sourlas, Phys. Rev.Lett. {\bf43}
(1979) 744.
\bibitem{Imbrie} J.Z.Imbrie, Phys. Rev.Lett. {\bf 43} (1984) 1747.
\bibitem{Bricmont} J.Bricmont and A. Kupiainen, Phys. Rev.Lett. {\bf 59}
(1987) 1829.
\bibitem{Kurchan} J. Kurchan, J. de Physique {\bf A1} (1992) 1333.
\bibitem{CDD} C. De Dominicis, Phys. Rev. {\bf B18} (1978) 4913
\bibitem{Mezard}M. M\'ezard and A. P. Young., Europhys. Lett.{\bf 18}
(1992) 653.
\bibitem{Martin}P.C. Martin, E.Siggia and H.Rose  Phys. Rev {\bf A8}(1973) 423.
\bibitem{PS} G.Parisi and N. Sourlas, Phys. Rev.Lett. {\bf 46} (1981) 871.
\bibitem{Zinn}J. Zinn-Justin "Quantum field theory and critical phenomena",
Oxford press, (1989).

\end{thebibliography}
\end{document}